\documentstyle[preprint,prl,aps]{revtex}
\tightenlines
\begin{document}
\draft
\title{Orbital current mode in elliptical quantum dots}
\author{Lloren\c{c} Serra, Antonio Puente}
\address{Departament de F\'{\i}sica, Universitat de les Illes Balears,
E-07071 Palma de Mallorca, Spain}
\author{Enrico Lipparini}
\address{Dipartimento di Fisica, Universit\`a di Trento,
and INFM sezione di Trento, 
I-38050 Povo, Italy}
\date{ August 31, 1999}
\maketitle
\begin{abstract}
An orbital current mode peculiar to deformed quantum dots is 
theoretically investigated; first by using a simple model that allows to 
interpret analytically 
its main characteristics, and second, by numerically solving the 
microscopic equations 
of time evolution after an initial perturbation within the time-dependent 
local-spin-density approximation. 
Results for different deformations and sizes are shown.
\end{abstract}
\narrowtext

The study of collective excitations in semiconductor quantum dots is 
currently attracting much interest. Recent experiments,
using resonant Raman scattering in two-dimensional GaAs-AlGaAs quantum 
dots \cite{Sch96,Str94},
have probed both charge-density and spin-density collective 
excitations, as well as single-particle excitations. From the 
theoretical side, charge-density excitations have been investigated 
since several years ago 
with different approaches such as Hartree \cite{Broi90}, Hartree-Fock 
\cite{Gud91} and density-functional theory \cite{Ser99}. In addition, 
a general scheme to describe spin-density excitations
of both longitudinal and transverse character was presented in Refs.\ 
\onlinecite{Ser99,Lip99}. 
All these theoretical calculations applied the well known 
Random-Phase Approximation in circularly symmetric
dots, for which the angular momentum selection rules can be exploited.

Experiments on deformed nanostructures are nowadays providing 
very interesting new pieces of information (for instance, 
in Ref.\ \cite{Aus99} on ellipsoidal deformations).
This has prompted, in Refs.\ \onlinecite{Kos98,Hi99,Pue99,Yan99}, the 
extension of theoretical approaches to the symmetry-unrestricted 
situation. 
Motivated by this exciting new direction of the quantum dot field we
report in this communication on a novel class of collective excitations. 
It
involves the generation of orbital currents and is peculiar to deformed
quantum dots. We will show that the orbital current excitation (OCE)
is strongly connected to the quadrupole charge-density excitation
(QCDE) and that its most clear signature
is in the magnetic dipole strength (M1). The characteristics of this 
mode 
will be investigated first analytically, by using a simplified model, 
and second, by solving numerically the time-dependent Kohn-Sham equations 
corresponding to this particular motion. It is worth to point out that 
current modes similar to this one have been measured \cite{Boh84}
and theoretically predicted \cite{Lip83} in atomic nuclei, and are 
also expected to exist in deformed metal clusters\cite{Lip89} and
in the condensate of trapped bosons.
In fact, while writing the present communication we have become aware 
of a microscopic
calculation for metal clusters, using an schematic 
Random-Phase Approximation, that has been 
published in Ref.\ \onlinecite{Nest99} and one for Bose gases which is 
in preparation \cite{Str99}.

Let us assume a perfectly elliptic quantum dot, whose electron $\rho$
and kinetic energy $\tau$ densities are functions of the ellipse 
contour lines
\begin{eqnarray}
\rho({\bf r}\,) &\equiv& \rho\left({x^2\over R_x^2} 
+{y^2\over R_y^2}\right)\nonumber\\
\tau({\bf r}\,) &\equiv& \tau\left({x^2\over R_x^2} +{y^2\over R_y^2}
\right)\; ,
\label{eq1}
\end{eqnarray}
where $R_x$ and $R_y$ are the ellipse radii. If $R_x=R_y$ we obviously 
recover the circularly symmetric densities. A time-dependent displacement 
may be represented as $\alpha(t){\bf u}({\bf r}\,)$, where $\alpha(t)$ is 
a time dependent parameter and ${\bf u}({\bf r}\,)$ gives the vector 
displacement at point ${\bf r}$. The corresponding displacement operator 
${\cal D}$, that acting on the $N$-electron ground state gives the 
displaced state $|d\rangle = {\cal D} |0\rangle$, is 
\begin{equation}
{\cal D} = \exp{\left[
{i\over 2\hbar} \alpha(t) \sum_{i=1}^N
{({\bf u}({\bf r}_i)\cdot{\bf p}_i + H.a.\ ) } \right] }\; ,
\label{eq2}
\end{equation}
where ${\bf p}_i$ is the $i$-th electron momentum operator and H.a.\ 
stands for the Hermitian adjoint operator.

The OCE is represented by the following displacement field
\begin{equation}
{\bf u}({\bf r})= \hat{\bf e}_z\times{\bf r}+
\eta\nabla(xy) \equiv
\left(
      -y(1-\eta)\, ,\,
      x(1+\eta)\right)\; .
\label{eq3}
\end{equation}
It is a combination of a rigid rotation with $z$ axis and a quadrupole 
distortion, weighted with a parameter $\eta$. This is a divergency free 
field $\nabla\cdot{\bf u}=0$ and, by choosing 
\begin{equation}
\eta={R_y^2-R_x^2\over R_x^2+R_y^2}\; ,
\end{equation}
it can be shown that the density variations up to second order in 
$\alpha$
are 
\begin{eqnarray}
\delta\rho &=& 0\nonumber\\
\delta\tau &=& {\alpha^2\over 6}\sum_{k\ell}(\nabla_k u_\ell + 
\nabla_\ell u_k)^2 \tau\; .
\end{eqnarray}
This result implies that the collective motion does not modify the 
electron
density and therefore, it is not affected by the Coulomb interaction. 
In fact the cost in the Coulomb energy is minimized by adding the
quadrupole term $\nabla(xy)$ in the
displacement field (\ref{eq3}).
The collective motion associated with this field 
modifies the kinetic energy density and thus it is an example of an
elastic (or shear) mode  
exhibited by a Fermi system, that have been extensively studied in atomic 
nuclei \cite{Lip83}. The electronic motion is in fact a collective 
rotating flow along the ellipse contour lines.

The frequency of the OCE may be estimated assuming the oscillator 
formula $\omega_{\rm OCE}=\sqrt{k\over M}$, where the restoring force
$k=2E({\bf u})$ is fixed by the energy change $E({\bf u})$ associated with
the displacement field (\ref{eq3}), 
$M=m\int {\bf u}^2 \rho d{\bf r}$ is the collective mass parameter
and $m$ is the single-electron effective mass.
One gets
\begin{eqnarray}
k &=& {16\over 3} \eta^2 E^{(0)}_{\em kin} \nonumber\\
M &=& N m \langle r^2\rangle (1-\eta^2)\; ,
\end{eqnarray}
where $E^{(0)}_{\em kin}$ is the unperturbed kinetic energy. If we 
further
assume $\langle r^2\rangle = {1\over 2} r_s^2 N$ and $E^{(0)}_{\em kin}=
{1\over 2} \varepsilon_F N$ (where $\varepsilon_F$ and $r_s$ are the 
Fermi
energy and the Wigner-Seitz radius, respectively) we finally get
\begin{equation}
\omega_{\rm OCE}\approx \sqrt{16\over 3} {\hbar\over m r_s^2} 
{\eta\over \sqrt{1-\eta^2}} N^{-1/2} \; .
\label{OCE}
\end{equation}
This simple expression tells us how the electronic OCE scales with $N$, 
deformation of the ellipse $\eta$, and electronic density ($r_s$ 
parameter).

Repeating a similar treatment with a pure quadrupole displacement 
${\bf u}=\nabla(xy)$ the frequency of the QCDE can be 
estimated
\begin{equation}
\omega_{\rm QCDE} \approx \sqrt{2}\omega_0\;,
\label{QCDE}
\end{equation}
where $\omega_0$ is the average parameter of the external 
confining potentials in $x$ and $y$ directions, assumed of parabolic 
type,
i.e., $\omega_0=(\omega_x+\omega_y)/2$ (see below).

From a physical point of view we expect both modes, OCE and QCDE, to 
manifest themselves in the response to the magnetic orbital dipole (M1) 
operator $\mu_B L_z$, and to the electric quadrupole (E2) operator $xy$. 
However, to really ascertain what is the relative weight of each mode in 
these channels we need to perform a more microscopic calculation. This is 
our goal in what follows.

We will describe the time evolution within density functional theory, 
in the local spin-density approximation. The time dependent 
Kohn-Sham equations are solved by discretizing the $xy$ plane in a 
uniform grid of equally spaced points and using the Crank-Nicholson 
approximation. 
Of course, the unperturbed state is the Kohn-Sham ground state, 
numerically obtained by solving the static Kohn-Sham equations in 
the same grid by a steepest descent method. Technical details of the 
method can be found in \cite{Pue99}. 

The set of single-particle orbitals $\{ \varphi_i({\bf r})\}$ evolves in 
time as \cite{units}
\begin{equation}
i{\partial\over\partial t} \varphi_{i\eta}({\bf r},t) =
h_\eta[\rho,m]\, \varphi_{i\eta}({\bf r},t)\;,  
\label{eqh}
\end{equation}
where the spin index is $\eta=\uparrow,\downarrow$, and total density 
and magnetization are given in terms of the spin densities 
$\rho_\eta({\bf r})=\sum_i{|\varphi_{i\eta}({\bf r})|^2}$, by 
$\rho=\rho_\uparrow+\rho_\downarrow$ and 
$m=\rho_\uparrow-\rho_\downarrow$, respectively. The Hamiltonian $h_\eta$ 
in equation (\ref{eqh}) contains, besides the kinetic energy, 
the confining potential 
$v^{({\em conf})}({\bf r})$, the Hartree potential 
$v^{(H)}({\bf r})=\int{d{\bf r}' \rho({\bf r}')/{|{\bf r}-{\bf r}'|}}$,
and the exchange-correlation piece
$v^{(xc)}_\eta({\bf r})=
{\partial\over\partial\rho_\eta}{\cal E}_{xc}(\rho,m)$.
The exchange-correlation energy density ${\cal E}_{xc}(\rho,m)$ has been 
described as in Refs.\ \cite{Kos98,Hi99,Pue99}.

To model the elliptic quantum dots we will follow the prescription of 
Refs.\ \cite{Aus99,Hi99} and consider the confinement produced by 
anisotropic parabolas with parameters $\omega_x$ and $\omega_y$ in $x$ 
and
$y$ directions, respectively. We define the ratio 
$\beta=\omega_y/\omega_x$
and fix the centroid with the Wigner-Seitz radius as 
$\omega_0^2=1/r_s^3\sqrt{N}$. 
In terms of these parameters, the external confining potential reads
\begin{equation}
v^{(\em conf)}({\bf r}) = {1\over 2} \omega_0^2
{4\over (1+\beta)^2}(x^2 + \beta^2 y^2)\; .
\end{equation}
As discussed in Ref.\ \cite{Aus99}, this is a reasonable approximation 
to the real confining potential in vertical quantum dots
with rectangular (mesa) structure.

An initial perturbation, modelling the interaction with the physical
probe, is needed in order to excite the system and monitor its time 
evolution. 
This is achieved by modifying the orbitals with the 
displacement operator (\ref{eq2}). With our previous discussion, 
three natural options for the displacement field ${\bf u}({\bf r})$ 
come inmediately to mind: a pure rotation (twist), a pure quadrupole
distortion and the combination given in Eq.\ (\ref{eq3}) (orbital 
distortion).
In the latter case, we fix $\eta$ with the elliptic potential 
parameters by assuming that $R_x/R_y=\beta $, i.e., the ratio between
$x$ and $y$ radii is given by the inverse ratio of the corresponding 
parabola coefficients. By looking at the density contour lines we 
have checked that this assumption is well satisfied. 
It yields
\begin{equation}
\eta={1-\beta^2\over 1+\beta^2}\;.
\end{equation}

Figure 1 shows the results for the $N=20$ electron dot confined with 
$r_s=1.51 a_0^*$ and $\beta=0.75$ 
(i.e., $\omega_x$=0.29~H$^*$, $\omega_y$=0.22~H$^*$). 
This figure nicely confirms the 
results anticipated with the analytical model. The orbital M1 strength
is divided into two clear regions. One at high energy, which is 
associated 
with the QCDE, and one at low energy associated with the OCE. The relative 
weigth given to both states is sensitively controlled with the parameter 
$\eta$. When this is forced to zero (twist), the QCDE takes a large part 
of the M1 strength. But when $\eta$ is taken according to the system 
deformation,  the OCE is the dominant mode. The 
decoupling 
is not perfect, as it is in the simple model, and the short period signal of 
the QCDE can be clearly seen, although with much lesser amplitude than 
for the twist excitation. The analytical expressions (\ref{OCE}) and 
(\ref{QCDE})
yield for this case the numerical values 0.066 H$^*$ and 
0.36 H$^*$, respectively, which agree quite reasonably with the 
microscopic result.
The OCE is strongly fragmented because of Landau damping. This can be 
seen from Fig.\ 1, since the OCE overlaps with peaks associated with 
single particle excitations, 
obtained by keeping $h_\eta$ fixed to its static value in Eq.\ 
(\ref{eqh}). By contrast, the QCDE is moved to higher energy by the 
residual interaction
and is not fragmented. The lower panels of Fig.\ 1 show the M1 signal 
after
an initial quadrupole distortion. In this case, the QCDE is the most 
dominant
and we have used a logarithmic vertical scale to show that the 
OCE intensity is around 2\% of the QCDE one. 
We conclude from Fig.\ 1 that the OCE produces a clear signature in the 
experimental orbital M1 strength of elliptical dots.  

We have peformed the same analysis with the $xy$ signal, corresponding
to the E2 channel, with similar conclusions, although the QCDE is more 
dominant in this case. The fragmentation patterns are the same as for 
the M1 results already discussed and the percentage of the OCE highest 
peak 
with respect to QCDE is $\sim 1$\%, $\sim 75$\% and $\sim 1$\% for the 
twist, orbital and quadrupole distortions, respectively.

The elastic behaviour of the OCE, in contrast to charge-density excitations, 
can be appreciated
from the time evolution of the different contributions to the total 
energy after the initial 
perturbation. As seen from Fig.\ 2, after a rigid twist the system 
increases only its energy in the external field. This triggers the motion 
and, as a consequence of total energy conservation, the other energy 
contributions begin to increase at the expense of the external field 
term.
With perturbation (\ref{eq3}) the situation is qualitatively different.
At $t=0$ there is an important increase in both kinetic and internal 
Coulomb energies. The first is due to the elasticity of Fermi systems 
mentioned before, while the second appears because the quantal densities 
are not really uniform within the ellipsoid, but have shell oscillations. 
The rotation along the elliptic contour deforms this structure and thus
increases the internal Coulomb energy. Again, because of energy 
conservation, once the motion has started all contributions fluctuate 
in a correlated manner. 

Figure 3 shows the evolution with deformation of the M1 strength, for 
$N=20$ and $r_s=1.51 a_0^*$, after an initial orbital displacement (\ref{eq3}). 
In each case, the positions of the analytical energies (\ref{OCE}) and 
(\ref{QCDE}) are indicated by arrows. 
The energy dependence of the OCE is in qualitative agreement with Eq.\ 
({\ref{OCE}), decreasing from 
$\beta=0.5$ to 0.875. However, formula (\ref{OCE}) begins to deviate from 
the miscrocopic result for the highest deformation ($\beta=0.5$).
The QCDE does not change much with deformation since we keep the centroid 
$\omega_0$ fixed.
It can also be seen from this figure that when $\beta$ approaches unity, 
i.e., the dot becomes circular, 
the OCE carries less strength with respect to the QCDE. In fact, for 
$\beta=0.875$ we have used logarithmic scale, since the OCE height
is only $\sim$1\% of the QCDE one.

In Fig.\ 4 we show the M1 strength for a fixed $\beta$ and 
varying the number of electrons. There is a tendency to decrease the 
OCE energy with increasing size, correctly pointed by the analytical 
expression (\ref{OCE}). It is also seen that the fragmentation of the OCE 
increases with size because of the increasing role of Landau damping. 

To summarize, the existence of an intense, low lying OCE in the M1 
strength of elliptical quantum dots has been pointed out. The importance 
of this mode 
with respect to the QCDE, as well as its evolution 
with dot deformation and size, have been discussed. 
Our analysis is based on the time 
evolution of the microscopic Kohn-Sham equations and on a simplified 
model that allows to obtain analytical solutions. The validity of the 
analytical expressions has been checked by comparison with the 
microscopic result. The microscopic calculation has also shown
a sizeable fragmentation of this low energy mode, that we attribute 
to an important role of Landau damping.

This work was supported in part by Grant No.\ PB95-0492 from CICYT, 
Spain.

\begin{figure}[h]
\caption{Results for the time evolution of an elliptic dot with $N=20$, 
$\beta=0.75$ and $r_s=1.51 a_0^*$, with the three different initial 
perturbations (rigid twist, orbital and quadrupole distortions).
Left panels display the simulated orbital M1 signal as a function of 
time while right panels show the corresponding strength
functions in arbitrary units. 
Middle right panel also shows the independent particle 
strength 
function (dashed line) and the position of the analytical 
approximations (\ref{OCE}) and (\ref{QCDE}) with arrows.}
\end{figure}

\begin{figure}[h]
\caption{Time dependence of the different energy variations after
an initial twist and orbital perturbations: kinetic (solid), 
internal Coulomb (dash), external confining field (dash-dot), 
exchange-correlation (dots) and total (dash-dot-dot).
Notice that the sum of kinetic, Coulomb, external field and 
exchange-correlation contributions yield the total energy increment 
and remains constant in time.}
\end{figure}

\begin{figure}[h]
\caption{Dependence of the OCE and QCDE with deformation
($\beta$ parameter, see text) for $N=20$ electrons and $r_s=1.51 a_0^*$.
The arrows indicate the energies given by Eqs.\
(\ref{OCE}) and (\ref{QCDE}).}
\end{figure}

\begin{figure}[h]
\caption{Dependence with size of the OCE and QCDE strength for a fixed 
deformation of $\beta=0.75$. As in Fig.\ 3, the arrows indicate the 
energies of the analytical expressions
(\ref{OCE}) and (\ref{QCDE}).}
\end{figure}


\begin{references}
\bibitem{units} We use effective atomic units corresponding to GaAs
values, i.e., H$^*\approx 12$ meV, $a_0^*\approx 98$ {\AA} and 
$\tau^*\approx 55$ fs, as energy, length and time units.
\end{references}
\end{document}